\documentclass[prd,a4paper,twocolumn,superscriptaddress, amsmath,amssymb,
aps,
mathtools,
enumitum]{revtex4-1}
\usepackage{graphicx}
\usepackage{epstopdf}
\usepackage{dcolumn}
\usepackage{bm}
\usepackage{hyperref}
\graphicspath{{./Figures/}}
\newcommand{\be}{\begin{equation}\begin{aligned}}
\newcommand{\ee}{\end{aligned}\end{equation}} 
\newcommand{\li}{\mathcal{L}_{\xi}}
\newcommand{\hmod}{H_{\text{mod}}}

\newcommand{\bL}{\mathbf{L}}
\newcommand{\bTheta}{\mathbf{\Theta}}
\newcommand{\bE}{\mathbf{E}}

\newcommand{\bJ}{\mathbf{J}}
\newcommand{\bQ}{\mathbf{Q}}

\newcommand{\bOmega}{\pmb{\Omega}}
\newcommand{\bC}{\mathbf{C}}

\begin{document}
	\preprint{}
	
	\title{Holographic equivalence between the first law of entanglement entropy and the linearized gravitational equations}
	
	\author{Benjamin Mosk}
	\affiliation{Stanford Institute for Theoretical Physics, Department of Physics, Stanford University\\
		Stanford, California 94305, USA}


\begin{abstract}
We use the intertwining properties of integral transformations to provide a compact proof of the holographic equivalence between the first law of entanglement entropy and the linearized gravitational equations, in the context of the AdS/CFT correspondence. We build upon the framework developed by Faulkner \emph{et al}. \cite{Faulkner:2013ica} using the the Wald formalism, and exploit the symmetries of the vacuum modular Hamiltonian of ball-shaped boundary regions. 
\end{abstract}

\maketitle

\section{Introduction and Summary} The AdS/CFT correspondence is the conjecture that under certain conditions, a theory of gravity in a $(d+1)$-dimensional asymptotically Anti de Sitter (AdS) spacetime (the bulk) is dual to a ``hologram'', a strongly coupled large-N conformal field theory (CFT) living on the $d$-dimensional asymptotic boundary \cite{Maldacena:1997re}. 

The AdS/CFT correspondence provides a framework for describing gravity in the bulk in terms of the CFT and may be \emph{the} way to describe --and learn about-- \emph{quantum} gravity. As a step in that direction, much effort has been put into ``deriving'' the classical gravitational laws from CFT properties and vice versa \cite{Allahbakhshi:2013rda,Nozaki:2013vta,Faulkner:2013ica,Swingle:2014uza}. 


It is hypothesized that spacetime itself can be seen as a geometrization of the \emph{entanglement structure} of the CFT \cite{VanRaamsdonk:2010pw}. For CFTs dual to Einstein gravity, the entanglement entropy of a boundary subregion $B$ is dual to the area of the bulk extremal surface $\tilde{B}$ that ends on --and is homological to-- $B$ (to leading order in 1/N) \cite{Ryu:2006bv,Hubeny:2007xt,Lewkowycz:2013nqa,Dong:2016hjy} 
\be\label{eq:hee}
S(B)_{\text{RT}} &= \frac{\text{Area}(\tilde{B})}{4G_N},
\ee
where $G_N$ is Newton's constant. We will refer to the extremal surface $\tilde{B}$ as the Ryu-Takayanagi, or RT-surface, also outside the realm of Einstein gravity. For a generalized theory of gravity (where the Lagrangian is a contraction of Riemann tensors), the entanglement entropy is thought to be dual the \emph{Wald-functional} evaluated at the extremal surface, up to terms involving the extrinsic curvature \cite{Hung:2011xb,Dong:2013qoa,Camps:2013zua}.

The holographic entanglement entropy (\ref{eq:hee}) thus furnishes a direct relation between properties of the CFT and the geometry of the bulk. This points to a relation between the ``dynamics'' of boundary entanglement and the dynamics of the bulk geometry: gravity. In \cite{Faulkner:2013ica} it was shown that the \emph{first law} of entanglement entropy, 
\be\label{eq:firstlaw}
 \delta \langle \hmod \rangle &= \delta S, 
\ee
where $\hmod$ is the modular Hamiltonian,
implies that the linearized gravitational equations must be satisfied in the bulk. 
The key ingredient in the derivation of \cite{Faulkner:2013ica} was that for ball-shaped boundary subregions $B$, there exists a $(d-1)-$form $\chi$ such that
\be\label{eq:chi}
\delta\hmod(B) &= \delta S(B) - \int_{\Sigma} d\chi, \ d\chi = \star (-2\delta E_{ab}\xi^b)
\ee
where the $\delta E_{ab}$ are the linearized gravitational equations of motion without matter coupling and $\Sigma$ is a Cauchy surface that ends on the RT-surface $\tilde{B}$ (see figure \ref{fig:xi}). The first law (\ref{eq:firstlaw}) now implies 
\be\label{eq:chivanish}
\int_{\Sigma(\tilde{B})}d\chi &= \int_{\Sigma(\tilde{B})}\star (-2\delta E_{ab}\xi^b) =0
\ee
Subsequently it is argued that the linearized gravitational equations must vanish locally by taking derivatives with respect to the ball radii.

In this article, we present a framework that combines the approach of \cite{Faulkner:2013ica} with methods in integral geometry \cite{Czech2016}. The key ingredient is to make optimal use of the symmetry properties of the vacuum modular Hamiltonian of a boundary ball, which satisfies two Casimir eigenvalue equations:
\be
\left(L^2_{\text{SO(d,2)}}+2d\right)\hmod = 0, \ \ \ \ \left(L^2_{\text{SO(d,1)}}+d \right)\hmod = 0.\nonumber
\ee
The second equation holds for all constant-time slices that can be associated to the spherical entangling surface $\partial B$. Both these Casimir eigenvalue equation operators ``annihilate'' the left-hand side of equation (\ref{eq:chi}) and thus provide a relation between $\delta S$ and the integral involving the linearized equations of motion. We will refer to these Casimir eigenvalue equation operators as \emph{Casimir equations} in what follows.

We show that the Casimir equations project the integral of $d\chi$ onto an integral of $\delta E_{ab}$ over the surface $\tilde{B}$. In particular, the SO(d,1) Casimir equation yields
\be\label{eq:deltaE}
0 &= \left(L^2_{\text{SO(d,1)}}+d\right)\int_{\Sigma}d\chi = 4\pi R\delta E_{tt}(\tilde{B}).
\ee
where $R$ is the \emph{Radon Transform}. The Radon transform is invertible on a constant-time slice \cite{helgason1959differential,Helgason1999}, and the fact that equation (\ref{eq:deltaE}) holds for all boundary balls on all constant-time slices then implies that the linearized gravitational equations must be satisfied. 

\begin{figure}
	\centering
	\includegraphics[width = 0.4\textwidth]{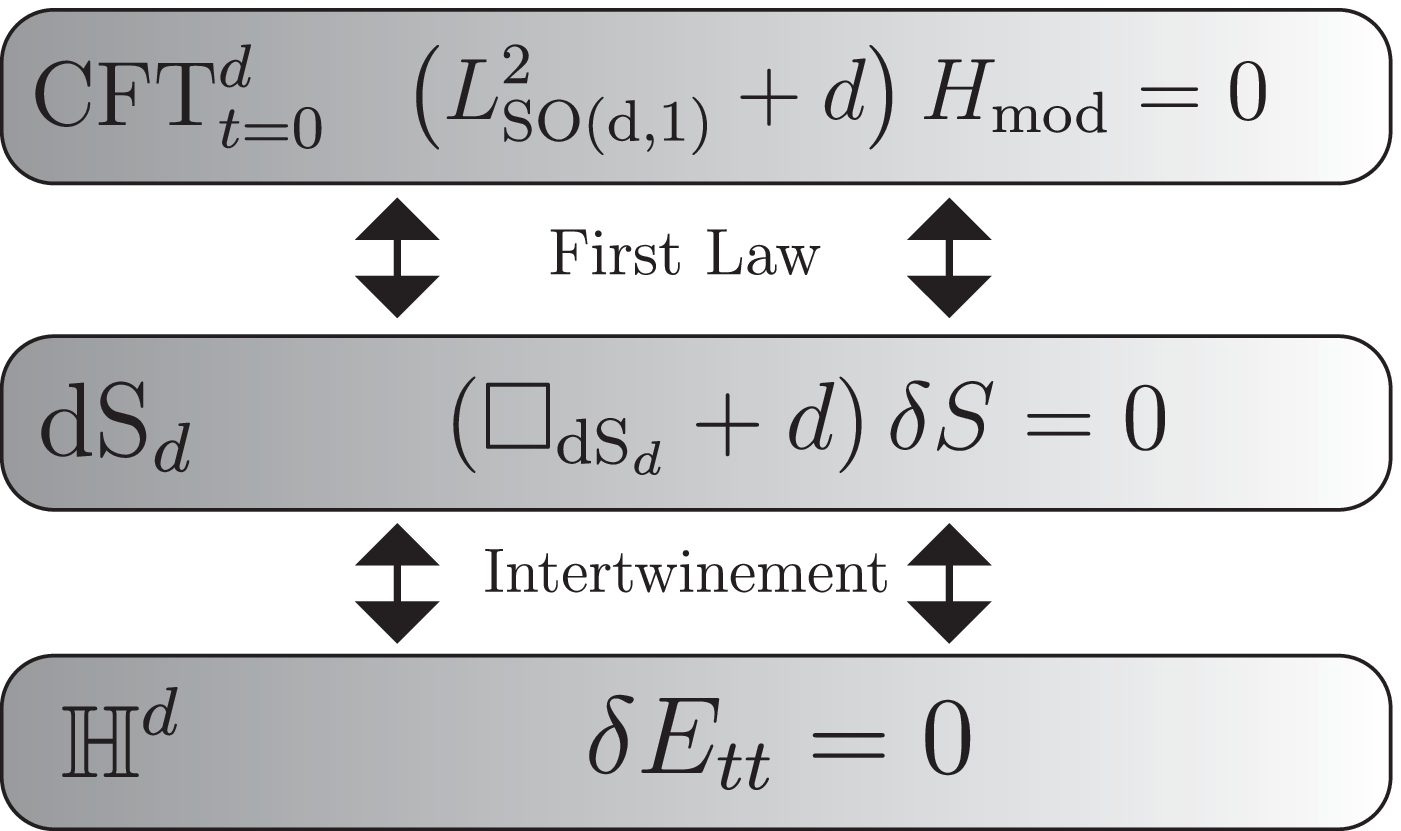}
	
	\caption{Summary}\label{fig:summary}

\end{figure}

This simple framework, summarized in figure (\ref{fig:summary}), that does not require a specific choice of coordinates or gauge, will be the subject of this article.  \\

\section{Wald Formalism}
Relation (\ref{eq:chi}) can be derived via the Wald formalism \cite{Wald:1993nt,LeeWald,Iyer:1994ys,Iyer:1995kg}, which we will briefly review. 

Consider a theory with Lagrangian $(d+1)$-form $\mathbf{L}$, a functional of the fields $\{\phi\}$, which variation is given by
\be\label{eq:varL}
\delta \mathbf{L} &= \mathbf{E}\left( \delta\phi\right) + d\bTheta (\delta\phi),
\ee
where $\mathbf{E}\left( \delta \phi\right)$ is the $(d+1)$-form containing the equations of motion and $\bTheta(\delta\phi)$ is called the symplectic potential, which appears as a boundary term in the action.

For a diffeomorphism invariant theory, the change of the Lagrangian under a diffeomorphism generated by a vector field $\xi$ is a total derivative
\be\label{eq:difL}
\delta \bL &= \li \bL = d \xi\cdot \bL +\xi \cdot d \bL = d \left(\xi\cdot \bL\right),
\ee
where the dot $\cdot$ stands for interior multiplication. The second equality is a manifestation of Cartan's formula. 

Equating (\ref{eq:varL}) and (\ref{eq:difL}) allows for the construction of a current $d$-form
\be\label{eq:current}
\bJ_{\xi} &= \bTheta (\li\phi)-\xi\cdot \bL, \text{ with }
&&d\bJ_{\xi} = -\bE (\li\phi).
\ee
The current $d$-form $\bJ_{\xi}$ is conserved \emph{on-shell}, for every diffeomorphism generating vector field $\xi$. As a consequence, there exists a $(d-1)$-form $\bQ_{\xi}$ \cite{Wald:1993nt}, called the ``Noether charge $(d-1)$-form'', such that, on-shell 
\be
\bJ_{\xi} &= d\bQ_{\xi}.
\ee We also define the \emph{symplectic current} $d$-form $\bOmega$
\be\label{eq:scurrent}
\bOmega (\delta_1\phi,\delta_2\phi) &= \delta_1 \bTheta(\delta_2\phi)-\delta_2 \bTheta(\delta_1\phi).
\ee
Using the equations above, it can be checked that
\be\label{eq:deltaj}
\delta \bJ_{\xi} 
&= \bOmega(\li\phi,\delta\phi)+d\left(\xi\cdot\bTheta(\delta\phi)\right) 
\ee
One can also define a Noether charge \emph{off-shell}
\be\label{eq:joffshell}
\bJ_{\xi} &= d \bQ_{\xi} + \xi^a \bC_a, \ \text{with }  \ d(\xi^a\bC_a) = -\bE(\li\phi)
\ee
where $\xi^a\bC_a$ is a $d$-form that contains the equations of motion for all the fields, except scalar fields \cite{Iyer:1995kg,Faulkner:2013ica} (see appendix \ref{app:noether} for a review). Note that $\bQ$ is not defined uniquely; we choose the standard definition in terms of the Wald functional \cite{Iyer:1994ys,Iyer:1995kg,Faulkner:2013ica}.

We proceed by considering a timelike Killing vector field $\xi$, with a bifurcation surface $\tilde{B}$. One can show that if a Hamiltonian $H_{\xi}$ can be constructed, that generates evolution along $\xi$, then, using equations (\ref{eq:deltaj}) and (\ref{eq:joffshell}), 
\begin{align}
\delta H_{\xi} &= \int_{\Sigma} \bOmega (\li\phi,\delta\phi) = \int_{\Sigma} \delta \bJ_{\xi} - \int_{\Sigma} d \xi\cdot \bTheta(\delta\phi),
\end{align}
where $\Sigma$ is a Cauchy surface extending from the bifurcation surface $\tilde{B}$ to the (asymptotic) boundary of the manifold under consideration. Using Stokes' Law we find 
\be\label{eq:mastereqn}
\delta H_{\xi} 
&= -\int_{\infty}\hspace{-2mm}\left(\delta\bQ_{\xi}-\xi\cdot \bTheta\right)+\int_{\tilde{B}}\hspace{-1mm}\left(\delta\bQ_{\xi}-\xi\cdot \bTheta\right)+\int_{\Sigma}\hspace{-1mm} \xi^a\delta\bC_a
\ee
This equation will form the basis for the remainder of this article. \\

\section{The Holography of Boundary Balls} 
The RT-surface $\tilde{B}$ for a ball-shaped boundary region $B$ is highly symmetric: it is the bifurcation surface of a Killing vector field $\xi(B)$ \cite{Faulkner:2013ica}. This observation sets the stage for a natural application of the Wald formalism, from which equation (\ref{eq:chi}) can be derived.

\begin{figure}
	\includegraphics[width = 0.4\textwidth]{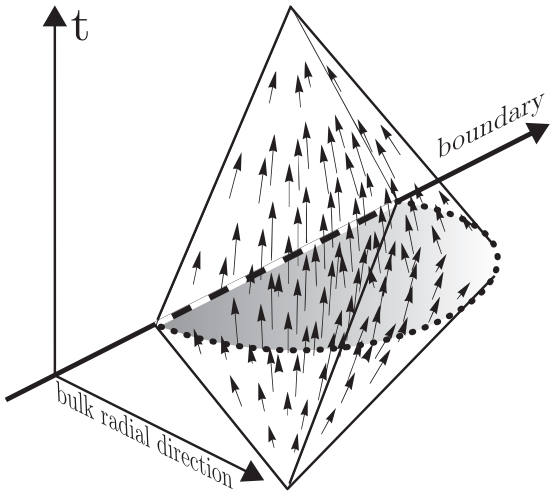}
	\caption{The Killing vector field $\xi$. The dashed line represents the boundary ball $B$ and the dotted line represents the RT-surface $\tilde{B}$. The shaded area is a Cauchy surface $\Sigma(\tilde{B})$ that lies on a constant-time slice.}\label{fig:xi}
\end{figure}

The entanglement entropy of ball-shaped boundary subregions is \emph{also} interesting for the following reasons:
\begin{itemize}
	\item the reduced density matrix $\rho_B$ is thermal with respect to the Hamiltonian $\hmod = -\log \rho_B$, which is the charge associated with the modular flow generating Killing vector field $\xi(B)$ \cite{Casini2011,Faulkner:2013ica}
	\item $\hmod(B)$ can be written as the integral of a smeared, local operator, the stress tensor $T$ \cite{Casini2011}:
	\be\label{eq:modham}
	\hmod(B) &=  \int_{B}  \star j, \ \ \ j_a = T_{ab}\xi^b
	\ee
	\item the Killing vector field $\xi(B)$ can be uniquely continued into the bulk causal wedge, the \emph{AdS-Rindler wedge} (see figure \ref{fig:xi}). We normalize $\xi(B)$ to have surface gravity $2\pi$.
\end{itemize}

The boundary stress tensor's bulk dual is a functional of the bulk metric (perturbation) \cite{Balasubramanian:1999re,deHaro:2000vlm,Skenderis:2000in}. An important observation in \cite{Faulkner:2013ica} is that the holographic dual of the modular Hamiltonian is given by the contribution from the asymptotic boundary to equation (\ref{eq:mastereqn}):
\be\label{eq:hmod}
\hmod(B) &=  \int_{\infty}\left(\delta\bQ_{\xi}-\xi\cdot \bTheta\right).
\ee
For surfaces \emph{with vanishing extrinsic curvature}, the holographic entanglement entropy is given by \cite{Hung:2011xb,Dong:2013qoa}:
\be\label{eq:deltas}
\delta S(B) &= \int_{\tilde{B}}\left(\delta\bQ_{\xi}-\xi\cdot \bTheta\right) =  \int_{\tilde{B}} \delta \bQ_{\xi}  
\ee
where we used that $\xi|_{\tilde{B}}=0$. These geometrical identifications set the stage for the translation between entanglement dynamics (\ref{eq:firstlaw}) and gravitational dynamics.

\section{The Casimir equation}
The quadratic Casimir of the conformal group SO(d,2), $L^2_{\text{SO(d,2)}}$, has eigen operators that are labeled by their dimension $\Delta$ and spin $l$. The eigenvalues are given by \cite{Dolan:2000ut}
\be\label{eq:caseig}
C^{\small{\Delta,l}}_{\text{SO(d,2)}} &= -\Delta(\Delta-d)-l(l+d-2).
\ee 
The modular Hamiltonian $\hmod$ is a smeared integral of the boundary stress tensor (\ref{eq:modham}), which has dimension $\Delta = d$ and spin $l=2$, so $\hmod$ satisfies 
\be\label{eq:cas}
\left(L^2_{\text{SO(d,2)}}+2d\right)\hmod &= 0.
\ee
The modular Hamiltonian also satisfies a second type of Casimir equation for every constant-time slice that contains $\partial B$. Constant-time slices on the boundary are spacelike surfaces whose non-degenerate, timelike normal vector is Killing. They naturally extend to the bulk, and by symmetry, the RT-surface must lie on the bulk extension of the boundary constant-time slice. Thus, from now on, the term ``constant-time slice'' will also refer to its extension into the bulk, which has hyperbolic geometry. The stabilizer of a constant-time slice is the subgroup $\text{Isom}\{\mathbb{H}^d\} = \text{SO(d,1)}\subset$ SO(d,2). 

On a constant-time slice with normal vector $t$, the modular Hamiltonian (\ref{eq:modham}) can be written as 
\be\label{eq:hmodslice}
\hmod &=  \int_{B} d^{d-1}x \ T_{tt}|\xi(B)|.
\ee
Note that $t^a=\frac{\xi^a}{|\xi|}$, except on the bifurcation surface $\tilde{B}$. The stress tensor component $T_{tt}$ transforms as a scalar under isometries that preserve the constant-time slice. The modular Hamiltonian thus also satisfies an SO(d,1) Casimir equation on each constant-time slice $\Sigma_t(B)$:
\be\label{eq:slicecas}
\left(L^2_{\text{SO(d,1)}}+d\right)\hmod(B)  = 0,
\ee
where the eigenvalue is given by equation (\ref{eq:caseig}) with $d$ replaced by $d-1$ and $l=0$. $T_{tt}$ carries dimension $\Delta$ under SO(d,1) transformations, as it inherits its scaling behavior from the full SO(d,2) group. \\

\section{Intertwinement}
The modular Hamiltonian $\hmod(B)$ can be seen as a function on the space of boundary balls, which we will call \emph{kinematic space} \cite{Czech:2015qta,Czech2016}. The space of boundary balls on a single constant-time slice has a \emph{de Sitter} (dS) geometry and the SO(d,1) Casimir is represented as the d'Alembertian $\square_{\text{dS}_d}$ and the Casimir equation (\ref{eq:slicecas}) can be written as a de Sitter wave equation: $(\square_{\text{dS}_d}+d)\hmod =0$ \cite{deBoer:2015kda,Czech2016}.

We will exploit the \emph{intertwining properties} of integral transformations, which relate the d'Alembertian $\square_{\text{dS}_d}$ on kinematic space to the d'Alembertian $\square_{\mathbb{H}^d}$ on the constant-time slice, in order to prove equation (\ref{eq:deltaE}).

The \emph{Radon transform} $R$ and a second useful integral transform $R_{\xi}$ of a function $f$ are given by:
\be\label{eq:radonxi}
Rf(\tilde{B}) &\equiv \int_{\tilde{B}}f, \ \ \ \ \ \ \ R_{\xi}f(\tilde{B}) &\equiv
\int_{\Sigma_t (\tilde{B})}f|\xi(B)|
\ee
where $\Sigma_t \subset \mathbb{H}^d$ is taken to be on a constant-time slice.
It is well-known that the Radon transform satisfies an intertwinement relation \cite{ref1,Helgason1999,Helgason2011,Czech2016}
\be
L^2_{\text{SO(d,1)}} \cdot Rf &= R L_{\text{SO(d,1)}}^2 \cdot f
\ee
where $L^2_{\text{SO(d,1)}}$ is the quadratic $\mathfrak{so}(d,1)$ Casimir, and the $\cdot$ denotes its action on the object on its right hand side. For (homogeneous) coset spaces $G/H$, the quadratic Casimir of a (semi-simple) Lie group $G$ is represented on functions by the Laplacian, up to an overall scaling \cite{pilch1984}. Both hyperbolic space $\mathbb{H}^d$ as well as its kinematic space dS$_{d}$ are coset spaces of $G=$SO(d,1) with H=SO(d) and H=SO(d-1,1) respectively. In coordinates we have (see appendix \ref{app:intertwinement} for a review of intertwinement)
\be\label{int1}
\square_{\text{dS}}Rf &= -R\square_{\mathbb{H}^d}f. 
\ee
The second type of transform (\ref{eq:radonxi}) has a similar intertwining property:
\be\label{eq:int2}
\square_{\text{dS}}R_{\xi}f &= -R_{\xi}\square_{\mathbb{H}^d}f. 
\ee
It follows from a double partial integration and the Killing property of $\xi$ that:
\be\label{eq:int3}
\left(L^2_{\text{SO(d,1)}}+d\right)R_{\xi}f &= -2\pi Rf.
\ee
Note that intertwinement rules can be used to translate the dynamics of local fields on AdS spacetime to the dynamics of fields on kinematic space. The AdS/CFT correspondence provides a natural set of non-local, diffeomorphism invariant bulk probes that have a definite dual in the CFT. CFT Wilson loops, OPE-blocks and the two-point functions of heavy operators can be associated to extremal bulk quantities \cite{Maldacena:1998im,Czech2016,Balasubramanian:1999zv}, in addition to aforementioned entanglement entropies of boundary subregions. These non-local probes can (perturbatively) be seen as integral transformations of bulk fields, which also have intertwining properties; the derivation of intertwinement merely relies on diffeormorphism invariance (see appendix \ref{app:intertwinement}). \\


\section{Gravity}
At last we will combine the geometrical analysis (\ref{eq:mastereqn}) and the intertwinement rules (\ref{eq:int2},\ref{eq:int3}) to derive the linearized gravitational equations.

At leading order in 1/N, we can ignore matter fields. In that case, we have $\bOmega(\li g,\delta g) = 0$, since per definition $\li g = 0$. As a consequence, the left-hand side of equation (\ref{eq:mastereqn}) vanishes, so we have
\be\label{eq:onlygrav}
\overbrace{\int_{\infty}\hspace{-0.1cm}\left(\delta\bQ_{\xi}-\xi\cdot \bTheta(\delta\phi)\right)}^{\hmod (B)} &= \overbrace{ \int_{\tilde{B}}\left(\delta\bQ_{\xi}-\xi\cdot \bTheta(\delta\phi)\right)}^{\delta S(B)}-\int_{\Sigma} d\chi,
\ee
where (see appendix \ref{app:noether} for a review) \cite{Faulkner:2013ica}:
\be\label{eq:cexp} d\chi &\equiv  \xi^a\delta C^g_a = \star (-2\delta E^g_{ab}\xi^b).
\ee
The linearized gravitational equations satisfy $\nabla^a\delta E_{ab}=0$ by virtue of the Noether identity \cite{Faulkner:2013ica}. Conservation of $\delta E^g_{ab}\xi^b$ implies that the integral of $d\chi$ does not depend on the choice of the Cauchy surface $\Sigma(\tilde{B})$, which we take to be on a constant-time slice:
\be\label{eq:toconst}
\int_{\Sigma(\tilde{B})}d\chi &= \int_{\Sigma_t(\tilde{B})}d\chi = -2\int_{\Sigma_t(\tilde{B})} \delta E_{tt}|\xi|.
\ee
In \cite{Faulkner:2013ica}, the first law (\ref{eq:firstlaw}) and equation (\ref{eq:onlygrav}) are used directly to argue that the integral of the $d$-form $d\chi$ must vanish.
Subsequently, an appropriate combination of kinematic space derivatives is taken to argue that the equations of motion must vanish locally: $\delta E_{ab} = 0$. 

Here, we take a different approach: the key ingredient is equation (\ref{eq:onlygrav}), whose left-hand side is annihilated by the Casimir equation (\ref{eq:slicecas}), such that
\be\label{eq:conversion}
\left(L^2_{\text{SO(d,1)}}+d\right)\delta S(B)&=
\left(L^2_{\text{SO(d,1)}}+d\right) \int_{\Sigma_t} d\chi.
\ee
From equation (\ref{eq:conversion}) we see directly that the Casimir equation ``projects'' the integral over the Cauchy slice $\Sigma_t(\tilde{B})$ (right hand side) onto an integral that \emph{only} has support on the RT-surface $\tilde{B}$ only (left-hand side). 
Concretely, we recognize that equation (\ref{eq:toconst}) is of the type (\ref{eq:radonxi}). Applying intertwinement rule (\ref{eq:int3}) directly gives 
\be\label{eq:dett}
0 =&\left(L^2_{\text{SO(d,1)}}+d\right) \int_{\Sigma_t} d\chi \overset{\text{Eqn. \ref{eq:int3}}}{=} 4\pi R\delta E_{tt}(\tilde{B}).
\ee
The Radon transform is known to be invertible on hyperbolic space \cite{helgason1959differential,Helgason1999}. Equation (\ref{eq:dett}) holds for every boundary ball on every constant time-slice, so in every point and for every timelike vector $t^a$ we have
\be\label{eq:ttoall}
\delta E_{ab}t^at^b &= 0 \ \Rightarrow &&\frac{d}{dt^a}\frac{d}{dt^b} \left( \delta E_{cd}t^ct^d\right) = 0.
\ee
We conclude that the symmetric part of $\delta E_{ab}$ must vanish, which is equivalent to the condition that the linearized equations of motion must be satisfied. 

In \cite{Czech2017}, the Wald formalism and equation (\ref{eq:chi}) are \emph{not} used. Instead, it is shown that for theories with $S = (4G_N)^{-1}A$ the first law (\ref{eq:firstlaw}) leads to the linearized Einstein equations by directly applying intertwinement rules to $\delta S$, which is the \emph{longitudinal Radon transform} $R^{\parallel}$ \cite{Czech2016} of the metric perturbation $\delta g_{ab}$. Intertwinement rules are developed for the traceless and trace parts of the metric perturbation $\delta g_{ab}$, showing that 
\be
\left(L^2_{\text{SO(d,1)}}+d\right)\delta S(B)&= 4\pi R\delta E_{tt}(\tilde{B}),
\ee
which, in our approach directly follows for \emph{generalized} theories of gravity from equations (\ref{eq:conversion}) and (\ref{eq:dett}). \\
\section{Bulk Matter}
From the perspective of entanglement entropy, the quantum corrections of the holographic entanglement entropy (\ref{eq:hee}) are, to first subleading order in 1/N, given by the FLM-formula \cite{Faulkner:2013ana,Jafferis:2015del}
\be\label{eq:flm}
S(B) &= S_{\text{RT}}(\tilde{B})+S_{\text{bulk}}(\Sigma(\tilde{B}))+S_{\text{Wald-like}}(\tilde{B}) 
\ee
where $S_{\text{bulk}}$ is the bulk entanglement entropy for the AdS-Rindler wedge (see figure \ref{fig:xi}). The terms at subleading order in $G_N$ conspire to \cite{Faulkner:2013ana,Jafferis:2015del}
\be\label{eq:hbulk}
\delta S_{\text{bulk}}(\Sigma(\tilde{B}))+\delta S_{\text{Wald-like}}(\tilde{B}) &=  \int_{\Sigma(\tilde{\text{B}})}\star j, \ j_a = \delta\langle T_{ab}\rangle \xi^b,\nonumber
\ee 
where $T_{ab}$ is the \emph{Hilbert} stress tensor, which appears on the right hand side of the gravitational equations. Note that $T_{ab}$ contains a contribution from the graviton. This term modifies equation (\ref{eq:deltas}) at first subleading order in 1/N and is of the form (\ref{eq:radonxi}), on a constant-time slice. This means that intertwinement rule (\ref{eq:int3}) can be used, such that equation (\ref{eq:dett}) becomes
\be
2\pi R\left(2\delta E_{tt}- \delta \langle T_{tt}\rangle \right)(\tilde{B}) = 0,
\ee
which implies, by virtue of the invertibility of the Radon transform on $\mathbb{H}^d$ and equation (\ref{eq:ttoall}), that the linearized equations of motion must also be satisfied in the presence of matter. 

Conversely, if we \emph{assume} the linearized gravitational equations with matter coupling, then the invertibility of the Radon transform fixes the 1/N correction to be of the form (\ref{eq:hbulk}), which was first shown for \emph{Einstein} gravity in \cite{Czech2017}. \\

\section{Outlook}
In the above, we used the constant-time slice Casimir equation (\ref{eq:slicecas}). We could have used the conformal Casimir equation (\ref{eq:cas}) to ``annihilate'' $\hmod$. \\

In the appendix \ref{app:fullcasimir} we show that \be\label{eq:cas2}
0&= \left(L^2_{\text{SO(d,2)}}+2d\right)\left(\int_{\Sigma}d\chi +\int_{\Sigma} \star (\delta\langle T_{ab}\rangle \xi^b)\right) \\ &=-2\pi R^{\perp}(2\delta E^g-\delta\langle T\rangle),
\ee 
where $R^{\perp}$ is the perpendicular Radon transform \cite{Czech2017}, the integral of the projection of a tensor to the two-dimensional normal plane of $\tilde{B}$. Unfortunately, the inversion and injectivity properties of $R^{\perp}$ are still unknown.

Finally, an interesting observation is that for other homogeneous spaces the same framework can be applied, \emph{if} the terms in equation (\ref{eq:mastereqn}) can be identified as $\delta H$ (\ref{eq:hmod}) and $\delta S$ (\ref{eq:deltas}), to establish the equivalence of the linearized gravitational equations and the first law (\ref{eq:firstlaw}). One can also apply the Casimir equation in conjunction with intertwinement rules to the terms up to \emph{second order} in the perturbation, among which the \emph{canonical energy} $\mathcal{E}$ \cite{Lashkari:2015hha,Beach:2016ocq}, in order to obtain gravitational equations at second order in the perturbation. This will be the subject of future work. 
\begin{center}
	\textbf{Acknowledgments}
\end{center}
	I would like to thank Leonard Susskind for useful discussions. I am very grateful for the support of Bart{\l}omiej Czech, Lampros Lamprou, Samuel McCandlish and James Sully, my collaborators on \cite{Czech2017}. I would like to thank Ben Freivogel and Laurens Kabir for reading an early version of this manuscript. I want to thank Aitor Lewkowycz for his feedback on the first version of this article. BM is supported by The Netherlands Organisation for Scientific Research (NWO).

\appendix
\begin{widetext}
\section{$\xi^a\bC_a$ and the Noether Identity}\label{app:noether}
This section is based on appendix \textbf{B} of \cite{Faulkner:2013ica}.
Under a diffeomorphism generated by a vector field $\xi$, the variation of the Lagrangian n-form is given by
\be
\li \bL &= \bE(\li\phi)+d\bTheta (\li\phi).
\ee
The equation of motion n-form $\bE(\li\phi)$ contains all the fields. For an $(r,s)$-tensor field $\phi$, the contribution to $\bE(\li\phi)$ is given by
\be\label{eq:liphi}
\bE(\li\phi) &= \star E_{a_1\cdots a_r}^{b^1\cdots b_s} (\li\phi)^{a_1\cdots a_r}_{b^1\cdots b_s} 
\ee
where $E_{a_1\cdots a_r}^{b^1\cdots b_s}$ is the ``equation of motion tensor''. For example, for the metric field $g$, and Einstein gravity we have
\be
E^g_{ab} &= \frac{1}{16\pi  G_N}\left(G_{ab}+g_{ab}\Lambda\right).
\ee

Expanding the Lie-derivate in equation (\ref{eq:liphi}) gives

\begin{align}
	(E^\phi)_{a_1\cdots a_r}^{b_1\cdots b_s}\, \li \phi^{a_1\cdots a_r}_{b_1\cdots b_s}
	&=  (E^\phi)_{a_1\cdots a_r}^{b_1\cdots b_s}\,\left(\xi^c  \nabla_c \phi^{a_1\cdots a_r}_{b_1\cdots b_s} - \sum_{i=1}^r  \phi^{a_1 \cdots c \cdots a_r}_{b_1\cdots b_s}(\nabla_c \xi^{a_i})
	+\sum_{i=1}^s  \phi^{a_1 \cdots  a_r}_{b_1\cdots c\cdots b_s}(\nabla_{b_i} \xi^{c})
	\right)\nonumber\\
	&=  \xi^c\left( (E^\phi)_{a_1\cdots a_r}^{b_1\cdots b_s} \nabla_c \phi^{a_1\cdots a_r}_{b_1\cdots b_s}\right) 
	\\
	&+ \xi^c\left(\sum_{i=1}^r \nabla_d\left( (E^\phi)_{a_1 \cdots  c \cdots a_r}^{b_1\cdots b_s}
	\phi^{a_1 \cdots d \cdots a_r}_{b_1\cdots b_s}\right) 
	-\sum_{i=1}^s \nabla_{b_i}\left( (E^\phi)_{a_1 \cdots  a_r}^{b_1\cdots b_s}
	\phi^{a_1 \cdots  a_r}_{b_1\cdots c\cdots b_s}\right)\right)\nonumber  \\&+ 
	\nabla_c \left( \sum_{i=1}^{s} E^{b_1\cdots c\cdots b_s}_{a_1\cdots a_r}\phi^{a_1\cdots a_r}_{b_1\cdots d\cdots b_s}\xi^d -\sum_{i=1}^{r} E^{b_1\cdots b_s}_{a_1\cdots d\cdots a_r}\phi^{a_1\cdots c\cdots a_r}_{b_1\cdots b_s}\xi^d \right)\label{eq:lieexpans}
\end{align}
Equation (\ref{eq:lieexpans}) holds for \emph{any} vector field $\xi$, so we must have
\be\label{eq:noether}
\sum_\phi \left( (E^\phi)_{a_1\cdots a_r}^{b_1\cdots b_s} \nabla_c \phi^{a_1\cdots a_r}_{b_1\cdots b_s} 
+  \sum_{i=1}^r \nabla_d\left( (E^\phi)_{a_1 \cdots  c \cdots a_r}^{b_1\cdots b_s}
\phi^{a_1 \cdots d \cdots a_r}_{b_1\cdots b_s}\right) 
- \sum_{i=1}^s \nabla_{b_i}\left( (E^\phi)_{a_1 \cdots  a_r}^{b_1\cdots b_s}
\phi^{a_1 \cdots  a_r}_{b_1\cdots c\cdots b_s}\right)\right) = 0 
\ee
This is the ``Noether identity''. 

Now consider a theory in which only the metric appears. The Noether Identity (\ref{eq:noether}) becomes
\be
\nabla_d \left(E^g_{a_1c}g^{a_1d}+E^g_{ca_2}g^{da_2}\right) &= 2\nabla^dE_{cd} = 0.
\ee
If we assume that the unperturbed equations of motion are satisfied, and we expand in the perturbation it follows that $\nabla^b \delta E_{ab}=0$. In other words, the first order perturbation of the equations of motion is a conserved symmetric two-tensor. 

Using equations (\ref{eq:liphi}), (\ref{eq:lieexpans}) and the Noether identity (\ref{eq:noether}) we conclude
\be
\bE(\li\phi) &= \star \nabla_c \left( \sum_{i=1}^{s} E^{b_1\cdots c\cdots b_s}_{a_1\cdots a_r}\phi^{a_1\cdots a_r}_{b_1\cdots d\cdots b_s}\xi^d -\sum_{i=1}^{r} E^{b_1\cdots b_s}_{a_1\cdots d\cdots a_r}\phi^{a_1\cdots c\cdots a_r}_{b_1\cdots b_s}\xi^d \right) \\
&= \star \star d \star F \ \ \ \text{with: }F^c \equiv \left( \sum_{i=1}^{s} E^{b_1\cdots c\cdots b_s}_{a_1\cdots a_r}\phi^{a_1\cdots a_r}_{b_1\cdots d\cdots b_s}\xi^d -\sum_{i=1}^{r} E^{b_1\cdots b_s}_{a_1\cdots d\cdots a_r}\phi^{a_1\cdots c\cdots a_r}_{b_1\cdots b_s}\xi^d \right) \\
&= d\star F \equiv -d\left(\xi^a\bC_a\right)
\ee
The $\bC$ are defined, to be consistent with \cite{Faulkner:2013ica}, as:
\be
\xi^a\bC_a &= (-)\star F = (-)\star \left( \sum_{i=1}^{s} E^{b_1\cdots c\cdots b_s}_{a_1\cdots a_r}\phi^{a_1\cdots a_r}_{b_1\cdots a\cdots b_s}\xi^a -\sum_{i=1}^{r} E^{b_1\cdots b_s}_{a_1\cdots a\cdots a_r}\phi^{a_1\cdots c\cdots a_r}_{b_1\cdots b_s}\xi^a \right) 
\ee

For a theory with \emph{only} the metric field, we have
\be
\bE(\li g) &= d\star F = - d\left(\xi^a\bC^g_a\right), \ \ \text{with: }F_b = 2\xi^aE^g_{ab}
\ee
Clearly, for the unperturbed metric, both sides vanish (on-shell).
At the linear level, we have
\be
F_b &= 2\xi^a\delta E^g_{ab}, \ \ \  \ 
\chi = \star (-2\xi^a\delta E_{ab}).
\ee

\section{Killing Vector $\xi(B)$}
For completeness, we give the expression of the Killing vector $\xi$ in Poincar\'{e} coordinates. When the caustics of the boundary ball are parametrized by $x_{1,2} = (t_0\pm R,\vec{x}_0)$, where $R$ corresponds to the radius of the boundary ball and $(t_0,\vec{x}_0)$ to the center, then
\be\label{raamsd}
\xi (B(x_1,x_2))
&=  2\pi \frac{R^2-(t-t_0)^2-(\vec{x}-\vec{x}_0)^2-z^2}{2R}\partial_t - 2\pi \frac{(t-t_0)\left((\vec{x}-\vec{x}_0)\partial_{\vec{x}}+z\partial_z\right)}{R}.
\ee
The Killing vector $\xi$ also can be expressed in terms of vectors on embedding space $\mathbb{R}^{d,2}$. Let $N_{1,2}$ be the embedding space null vectors ``pointing towards'' the points $x_{1,2}$ on the asymptotic boundary of the AdS-hyperbola defined by $X^2=-1$. The Killing vector $\xi$ is now given by:
\be\label{eq:kvfemb}
\xi^A &= \frac{(N_2\cdot X)N_1^A-(N_1\cdot X)N_2^A}{N_1\cdot N_2}.
\ee
A more general expression in Poincar\'{e} coordinates, in terms of the boundary points $x_{1,2}$ is given by
\be\label{realcomp}
\xi^z &= z\frac{(z^2+(x-x_1)^2)- (z^2+(x-x_2)^2)}{(x_1-x_2)^2}, \ \ \ \ 
\xi^{\mu}
= \frac{(z^2+(x-x_1)^2)(x^{\mu}-x^{\mu}_2)-(z^2+(x-x_2)^2)(x^{\mu}-x^{\mu}_1)}{(x_1-x_2)^2}.
\ee

\section{Intertwinement}\label{app:intertwinement}
\subsection{Review of Intertwinement}
Under a diffeomorphism $x\mapsto x'(x)$ (that leaves the constant-time slice invariant) we have
\be
Rf(\tilde{B}) &\rightarrow Rf'(\tilde{B}') = Rf(\tilde{B}),
\ee
or in terms of the group element $g\in SO(d,1)$
\be\label{eq:gel}
Rg\cdot f (g\cdot \tilde{B}) &= Rf(\tilde{B}),
\ee
where $\cdot$ denotes the action of the group element on the object on its right hand side (function, surface,...). Equivalently to equation (\ref{eq:gel}), we can write
\be\label{eq:gelinv}
Rg\cdot f (\tilde{B}) &= Rf(g^{-1}\cdot\tilde{B}).
\ee
Since the isometry group $\text{Iso}\{\mathbb{H}^d\} = \text{SO(d,1)}$ is a Lie group, we can also write equation (\ref{eq:gelinv}) in infinitesimal form in terms of the generators of $\mathfrak{so}(d,1)$:
\be\label{eq:alg}
RL_{AB}\cdot f (\tilde{B}) &= -Rf(L_{AB}\cdot\tilde{B}),
\ee
where now the $\cdot$ denotes the action of the Lie-algebra element $L_{AB}$ on $C_0(\mathbb{H}^d)$ and $C(\mathcal{K}_{t=0})$ respectively. Exploiting relation (\ref{eq:alg}) twice, we find 
\be\label{eq:casimirs}
R L^2 \cdot f (\tilde{B}) &= Rf(L^2\cdot\tilde{B}) = L^2\cdot Rf(\tilde{B}),
\ee
where $L^2$ is the quadratic Casimir operator. For homogeneous coset spaces G/H, the quadratic Casimir of the (semi-simple) Lie group G is represented by the d'Alembertian \cite{pilch1984}. Both $\mathbb{H}^d$ as well as the kinematic space Isom$\{\mathbb{H}^d\}=$dS$_d$ are of the form G/H, with G=SO(d,1) and H=SO(d) and H=SO(d-1,1) respectively. The \emph{relative} scaling is fixed by considering the Cartan Killing form on $\mathfrak{so}(d,1)$ explicitly, or simply by  checking the intertwinement property in a coordinate basis. One can check that equation (\ref{eq:casimirs}) becomes
\be
\square_{\text{dS}_d} Rf(\tilde{B}) &= -R\square_{\mathbb{H}^d}f (\tilde{B}).
\ee
Nowhere did we use specific properties of $R$, so for 
\be
R_{\xi}f(\tilde{B}) &= -\int_{\Sigma_t(\tilde{B})}f |\xi(\tilde{B})|
\ \ \text{   we have similarly:   } \ \ \ \
\square_{\text{dS}_d}R_{\xi}f(\tilde{B}) = -R_{\xi}\square_{\mathbb{H}^d}f (\tilde{B}).
\ee
\subsection{Intertwinement for $R_{\xi}$}\label{eq:int3der}
In this subsection we show that for a \emph{conserved} symmetric two-tensor $W$
\be
\left(L^2_{\text{SO(d,1)}}+d\right) \int_{\Sigma(\tilde{B})} \star (W_{ab}\xi^b) &= -\int_{\tilde{B}} W_{ab}\frac{\xi^a\xi^b}{|\xi|^2},
\ee
where $\tilde{B}$ is the bifurcation surface for the Killing vector $\xi$, and $\Sigma$ is any Cauchy surface that ends on $\tilde{B}$ and $B$.
This is the more general form of intertwinement property (\ref{eq:int3}) for (\ref{eq:radonxi}). For completeness, we will also derive relation (\ref{eq:int3}) here, \emph{not} using any specific coordinate set. 

First, we note that the integral does not depend on the choice of $\Sigma(\tilde{B})$, by virtue of the conservation of $W_{ab}\xi^b$:
\be
\nabla^a (W_{ab}\xi^b) &= (\nabla^aW_{ab})\xi^b + W_{ab}\nabla^a\xi^b = 0,
\ee 
by virtue of the conservation of $W$ and the Killing equation; we choose $\Sigma$ to be a surface orthogonal to $\xi$:
\be
\int_{\Sigma(\tilde{B})} \star (W_{ab}\xi^b) &= \int_{\Sigma_t(\tilde{B})}  \left(W_{ab}\frac{\xi^a\xi^b}{|\xi|}\right)|\xi| \equiv \int_{\Sigma_t(\tilde{B})}  f|\xi| \equiv R_{\xi}f(\tilde{B})
\ee
From the intertwinement property (\ref{eq:int2}) we have
\be
L^2_{\text{SO(d,1)}} \int_{\Sigma_t(\tilde{B})} f|\xi| &= -\int_{\tilde{B}} (D_aD^af)|\xi|,
\ee
where $D$ is the induced covariant derivative on the $\mathbb{H}^d$. We can further simplify this result (\ref{eq:int2}) by using
\be\label{eq:exp}
(D_aD^af)|\xi| &=D_a\left(|\xi|D^af\right) -D_a\left(fD^a |\xi|\right) + fD_aD^a |\xi|,\nonumber 
\ee
One can check that
\begin{align}
	D_aD^a |\xi| &= \left(g^{ab}+\frac{\xi^a\xi^b}{|\xi|^2}\right)\nabla_a\nabla_b \sqrt{\xi^2} = \left(g^{ab}+\frac{\xi^a\xi^b}{|\xi|^2}\right)\nabla_a\left( \frac{\xi^c\nabla_b \xi_c}{\sqrt{\xi^2}}\right) \\
	&= \left(g^{ab}+\frac{\xi^a\xi^b}{|\xi|^2}\right)\left( \frac{(\nabla_a\xi^c)(\nabla_b \xi_c)}{\sqrt{\xi^2}}\right) -\left(g^{ab}+\frac{\xi^a\xi^b}{|\xi|^2}\right)\left(\frac{\xi^d\xi^c(\nabla_b \xi_c)(\nabla_a\xi_d)}{(\sqrt{\xi^2})^3}\right)\label{eq:k2} \\
	&+ \left(g^{ab}+\frac{\xi^a\xi^b}{|\xi|^2}\right)\frac{\xi^c\nabla_a\nabla_b \xi_c}{\sqrt{\xi^2}} \label{eq:k3}
\end{align}
Now for terms in (\ref{eq:k2}) we use that a surface orthogonal to $\xi$ has vanishing extrinsic curvature
\be
K_{ab} &= \left(g^{ac}+\frac{\xi^a\xi^c}{|\xi|^2}\right)\nabla_c\xi_b = 0
\ee
and for the last term (\ref{eq:k3}) we use that for Killing vectors on AdS-spacetime 
\be
\nabla_a\nabla_b\xi_c &= R_{cbad}\xi^d \overset{\text{AdS}}{=} -g_{ac}\xi_b+\xi_cg_{ab},
\ee
such that
\be\label{eq:boxxi}
\square_{\Sigma}|\xi|&= D_aD^a|\xi| = d|\xi|.
\ee
It follows that from equation (\ref{eq:exp}) and (\ref{eq:boxxi})
\be
\int_{\Sigma} (\square_{\mathbb{H}^d}f)|\xi| &= d \int_{\Sigma} f |\xi| - \int_{\tilde{B}} f N^aD_a|\xi| + \text{boundary terms at }\infty
\ee
We assume the other boundary terms to vanish, for sufficiently rapidly falling off $f$ at $\infty$ and using $\xi|_{\tilde{B}}=0$. In summary, we have used partial integration (\ref{eq:exp}) and equation (\ref{eq:boxxi}) to get
\be
\left(L^2_{\text{SO(d,1)}}+d\right) \int_{\Sigma(\tilde{B})} f|\xi| &= \int_{\tilde{B}} f N^aD_a|\xi|.
\ee
On the surface $\tilde{B}$, we have $\nabla_a\xi_b = \kappa n_{ab}$ where $n_{ab}$ is the anti-symmetric binormal and the surface gravity $\kappa=2\pi$. Since we integrated on a surface orthogonal to $\xi$, the normal vector $N$ is orthogonal to $\xi$ as well:
\be
N^aD_a|\xi| &= N^a\frac{\xi^c}{|\xi|}\nabla_a\xi_c =  2\pi N^a\frac{\xi^c}{|\xi|}n_{ac} = -\pi n^{ab}n_{ab} =  -2\pi\nonumber 
\ee
So finally we have,
\be
\left(L^2_{\text{SO(d,1)}}+d\right) \int_{\Sigma(\tilde{B})} f|\xi| &= \int_{\tilde{B}} f N^aD_a|\xi| = -2\pi \int_{\tilde{B}}f,
\ee
or
\be
\left(L^2_{\text{SO(d,1)}}+d\right) \int_{\Sigma(\tilde{B})} \star (W_{ab}\xi^b) &= -2\pi \int_{\tilde{B}} W_{ab}\frac{\xi^a\xi^b}{|\xi|^2}.
\ee

\section{The SO(d,2) Casimir Equation}\label{app:fullcasimir}
Here we prove equation (\ref{eq:cas2}). First we note that equation (\ref{eq:casimirs}) follows from diffeomorphism invariance only; specific details of the transformation $R$, the field $f$ and the isomorphism group were not used. So similar to equation (\ref{eq:casimirs}), we also have
\be
R L^2_{\text{SO(d,2)}} \cdot f (\tilde{B}) &=L^2_{\text{SO(d,2)}}\cdot Rf(\tilde{B}),
\ee
for any diffeomorhism invariant transformation $R$ of the field $f$, which can also be a tensor field.

Now we consider a particular transform $\tilde{R}_{\xi}$ that maps a conserved symmetric two-tensors $W_{ab}$ on AdS to a function on kinematic space:
\be\label{eq:rtilde}
\tilde{R}_{\xi}(\tilde{B}) &\equiv \int_{\Sigma(\tilde{B})} \star (W_{ab}\xi^b(\tilde{B}))
\ee
where $\Sigma(\tilde{B})$ is a Cauchy surface that ends on $\tilde{B}$. Note that (\ref{eq:rtilde}) does not depend on the specific choice of Cauchy surface by virtue of the conservation of $W_{ab}\xi^b(\tilde{B})$. Both the integral of $d\chi$ (see \ref{eq:chivanish}) as well as the FLM-formula (\ref{eq:flm}) are of this form. Below, we will derive the intertwining properties of the transform (\ref{eq:rtilde}).

On general tensors, the conformal Casimir $L^2_{\text{SO(d,2)}}$ is represented as \cite{Hijano:2015zsa,pilch1984}:
\be
-\square_{\text{AdS}}-l(l+d-1),
\ee
where $l=0$ for functions and $l=2$ for the traceless symmetric part of a two-tensor. Thus, decomposing $W_{ab}$ in its trace and traceless components
\be\label{eq:trace}
W_{ab}^{\text{trace}} &= \frac{W}{d+1}g_{ab}, \ \ \ \
W = g^{ab} W_{ab}, \ \ \ \
W_{ab}^{\text{traceless}} = W_{ab}-W_{ab}^{\text{trace}} 
\ee
we have

	\be
	(L^2_{\text{SO(d,2)}}+2d)\tilde{R}_{\xi}W &= \tilde{R}_{\xi} \left(-(\square_{\text{AdS}}W) + 2d W-2(d+1)W^{\text{traceless}}\right) \label{eq:ncurrent}
	\ee
	It follows after some algebra that
	\be\label{eq:hodgel}
	\tilde{R}_{\xi} \left(-(\square_{\text{AdS}}W) + 2d W-2(d+1)W^{\text{traceless}}\right) &=\int_{\Sigma(\tilde{B})} \star \left(\Delta (W_{ab}\xi^b) + 2 \nabla^b \mathcal{L}_{\xi}W_{ab}\right),
	\ee
	where $\Delta \equiv \delta d + d \delta$ is the Hodge Laplacian (and $\delta$ is the co-differential).

Two important identities used for the derivation of equation (\ref{eq:hodgel}) are given by:
\be
\left(\Delta \omega\right)_a &= -\left(\square \omega\right)_a + R_{a}^b\omega_b, \ \text{for a one form }\omega 
\ee
where $\square = \nabla^a\nabla_a$ and
\be\label{rule3}
\nabla^b \nabla_c W_{ab} &= \left[\nabla_b, \nabla_c \right]W_{ad}g^{bd} =  -(d+1)W_{ab}^{\text{traceless}}.
\ee
The second contribution to the RHS of equation (\ref{eq:hodgel}) vanishes, because under a diffeomorphism generated by $\xi$,
\begin{align}
	W_{ab} &\rightarrow \tilde{W}_{ab} &&= W_{ab}+(\mathcal{L}_{\xi}W)_{ab}+\dots &&& \\
	\nabla^bW_{ab} &\rightarrow \tilde{\nabla}^b\tilde{W}_{ab} &&= \nabla^b \tilde{W}_{ab} 
	&&&= \nabla^b \left(W_{ab}+(\mathcal{L}_{\xi}W)_{ab}+\dots\right) \label{eq:Wvanish}
\end{align}
where we use that $\xi$ is Killing, which implies that  $\nabla = \tilde{\nabla}$. Conservation of $W$ requires $\nabla^bW_{ab} = \tilde{\nabla}^b\tilde{W}_{ab} = 0$, so it follows from equation (\ref{eq:Wvanish}) that $\nabla^b \mathcal{L}_{\xi}W_{ab}=0$. In summary, we have
\be
\left(L^2_{\text{SO(d,2)}}+2d\right)\tilde{R}_{\xi}W &= 
\int_{\Sigma(\tilde{B})}\star \Delta (W_{ab}\xi^b). 
\ee
After some manipulation, using the definition of the Hodge Laplacian and Stokes theorem, it follows that

	\be
	\left(L^2_{\text{SO(d,2)}}+2d\right)\tilde{R}_{\xi}W &= -(-1)^{d+1} \int_{\Sigma(\tilde{B})}\star \star d \star d (W_{ab}\xi^b) &&\text{Note: }\delta (W_{ab}\xi^b) = (-1)^d\nabla^a(W_{ab}\xi^b)=0\\
	&= -\int_{\tilde{B}} \star d(W_{ab}\xi^b) &&\text{Note: }\star \star \omega_p = -(-1)^{p(n-p)}\omega_p  \\
	&=  -\int_{\tilde{B}} \star \left( W_b^c\nabla_a\xi_c-W_a^c\nabla_b\xi_c\right) &&\text{Note: }\xi|_{\tilde{B}}=0\\
	&= -2\int_{\tilde{B}} n^{ab} W_{[b}^c\nabla_{a]}\xi_c, &&
	\ee
where $n$ is the antisymmetric binormal.
On $\tilde{B}$,
$
\nabla_a\xi_b= \kappa n_{ab} = 2\pi n_{ab}
$, so it follows that
\be\label{eq:result}
\left(L^2_{\text{SO(d,2)}}+2d\right)\tilde{R}_{\xi}W &=2\pi\int_{\tilde{B}} s^{ab}W_{ab} \equiv 2\pi R^{\perp}W,
\ee
where $s^{ab}=(g^{ab}-h^{ab})$ is the symmetric binormal and $R^{\perp}$ is the \emph{transverse Radon transform} \cite{Czech2017}. Both $\delta E_{ab}$ and $\delta T_{ab}$ are conserved symmetric two-tensors, so
\be
0 &\overset{\text{Eqn. \ref{eq:cas}}}{=} &&\left(L^2_{\text{SO(d,2)}}+2d\right)\hmod(B) \\
&\overset{1^{\text{st}} \text{ law}}{=} &&\left(L^2_{\text{SO(d,2)}}+2d\right)\delta S(B) \\
&\overset{\text{Eqn. \ref{eq:conversion}}}{=} &&\left(L^2_{\text{SO(d,2)}}+2d\right)\tilde{R}_{\xi}\left(-2\delta E + \delta T\right) \\
&\overset{\text{Eqn. \ref{eq:result}}}{=} &&-2\pi R^{\perp}\left(2\delta E-\delta T\right)
\ee
This finalizes the proof of equation (\ref{eq:cas2}). This was first shown for Einstein gravity via a different method in (\cite{Czech2017}). Our result holds for a generalized theory of gravity.
\end{widetext}

\bibliography{bibliography}
\end{document}